# A Markov Chain approach to determine the optimal performance period and bad definition for credit scorecard

**Murphy Choy,**
School of Information System,
SMU, Singapore

**Ma Nang Laik,**
Assistant Professor,
School of Information System,
SMU, Singapore

## ABSTRACT

Performance period determination and bad definition for credit scorecard has been a mix of fortune for the typical data modeler. The lack of literature on these matters led to a proliferation of approaches and techniques to solve the problems. However, the most commonly accepted approach involves subjective interpretations of the performance period and bad definition as well as being chicken and egg problem. These complications result in poorly developed credit scorecard with minimal benefits to the banks. In this paper, we will be recommending a simple and effective approach to resolve these issues.

## INTRODUCTION

Credit risk scorecard is an important tool in the tool box of the banking industry. It has been widely used to control consumer credit risk and has been extended to small business credit risk (Anderson, 2005; Thomas et. al. 2002). The earliest credit scorecards were developed by Credit Scoring Consultancies as a way for finance companies to identify risky customers that should not have been given a loan. Due to their proprietary nature (or aptly statistical nature) (Anderson, 2005), few understood the mechanism of the scorecard at the point in time. Early practitioners of credit risk scorecard modeling spent massive amount of time refining the techniques used to build the scorecards. Besides refining the techniques, they spent a lot of time explaining the mechanism and philosophical approach to the finance companies to convince them to use the tool.

As time passes, more and more people understood the mechanism of the credit scorecard and are willing to adopt the model to manage their business. The sudden rise in the consumer credit market directly led to the rise of the credit scorecard industry marking a new milestone in the industry (Lewis, 1992). Many big credit scorecard consultancies were established during this period of expansion such as FICO and Experian which results in the huge disparity in the approaches taken to quantify the risk. This huge disparity results in a major argument about the philosophical aspect of credit scoring and how it should be applied.

At the beginning of the credit scorecard industry, they face strong opposition from a variety of established credit risk practitioner where they adopt conservative credit underwriting process which has been the traditional approach in the field. The main criticism against credit scorecard then was that the variables have very little relation to variables which models them and that the definition used in the modeling can be rather haphazard and offers little help to finance companies who are trying to manage these risks. This strong opposition is also voiced by some authors (Capon, 1976; Rosenberg et. al., 1994). While there has been much refinement of the credit scoring techniques in the banking world (Eisenbeis, 1977; Eisenbeis, 1978), many criticisms have not been satisfactorily resolved.

With the advent of Basel II, there has been widespread discussion about the definition of a bad account in the context of credit portfolio. The accepted definition for Basel II is any accounts with an ever 90 plus days past due within a performance period of 12 months is considered to be a bad account. This definition is controversial as different financial products behave differently. Some credit products such as mortgage takes a long time to any accounts to satisfy the bad definition while in other cases, the period is too long and most accounts will be considered bad by then (Thomas, 2002; Siddiqi, 2006). Thus proper definition is critical to both proper management of risk as well as operational needs of the banks. In the next section, we will describe the process of defining a bad definition and explore some of existing techniques in evaluating the most optimal combination for defining the performance period as well as the selected bad definition.

## DEFINING THE PROBLEM

Credit risk scorecard are designed to measure the probability of an event happening. To be able to measure such events, one must define the event in a manner that is easy to measure and does not confuse with other events that may be a combination of the events. The earliest credit scorecards have extremely simple target events such as predicting whether a customer will becomes ever 30 days past due in the next six months. The improvement in the raw computing power has resulted in ease of building more complicated models which attempts to capture more variations of the bad than what is traditionally used in modeling. Below are some examples of good definitions of 'bad' accounts and contrasting them with complicated and infeasible definitions.

| Bad Definition for Modeling | |
|---|---|
| **Good** | **Bad** |
| Ever X+ DPD in 3 Months | 2 Times X+ DPD in 4 Months |
| Ever 30+ DPD in 6 Months | 6 Times 30+ DPD in 12 Months |
| Ever 60+ DPD in 9 Months | 2 Times 30+ DPD and 4 Times X+ DPD in 10 Months |
| Ever 90+ DPD in 12 Months | 2 Times Consecutive 30+ DPD in 12 Months |

**Table 1: Bad definitions**

The problem with more complicated bad definition is the difficulty in truly understanding the outcome. Let us contrast the good and bad definitions and use the row 3 definitions from table 1. If you were to ask an analyst what it takes to be a bad customer, the answer will be the definition and you wonder, what about customers who are 1 times 90+ DPD or 3 times 30+ DPD in 10 months? Another possible situation that might arise from this definition is the simplification of the complex definition. The first condition is an extension to the second condition which implies that we can simplify slightly to '2 Times 30+ DPD and 2 Times X+ DPD in 10 Months'. One severe issue with using this type of definition is the time period needed. Given 6 times delinquent in 10 months, the probability of such an event will be very unlikely, resulting a small target population for modeling.

We have examples of good 'bad' definitions but we do not know which definition will meet the requirements of modeling credit default events. Getting a good definition for modeling both in terms of delinquency and performance period will be the focus of the next few sections. We will first discuss about the traditional approach of estimating the performance period and delinquency status for default prediction. Once we have discussed the weakness of the techniques, we will demonstrate the simplicity of the Markov chain approach which solves both problems simultaneously.

**EVER DPD CURVES ANALYSIS: PERFORMANCE PERIOD PROBLEM**

Determination of the performance period is typically achieved using a type of analysis called ever delinquency curves analysis. This analysis works by analysing the ever delinquency curves trend and attempts to identify the point where the rate of increment in the delinquency rates actually slows. Typically, this is done for several vintages for a particular delinquency. Below is an example of such a chart.

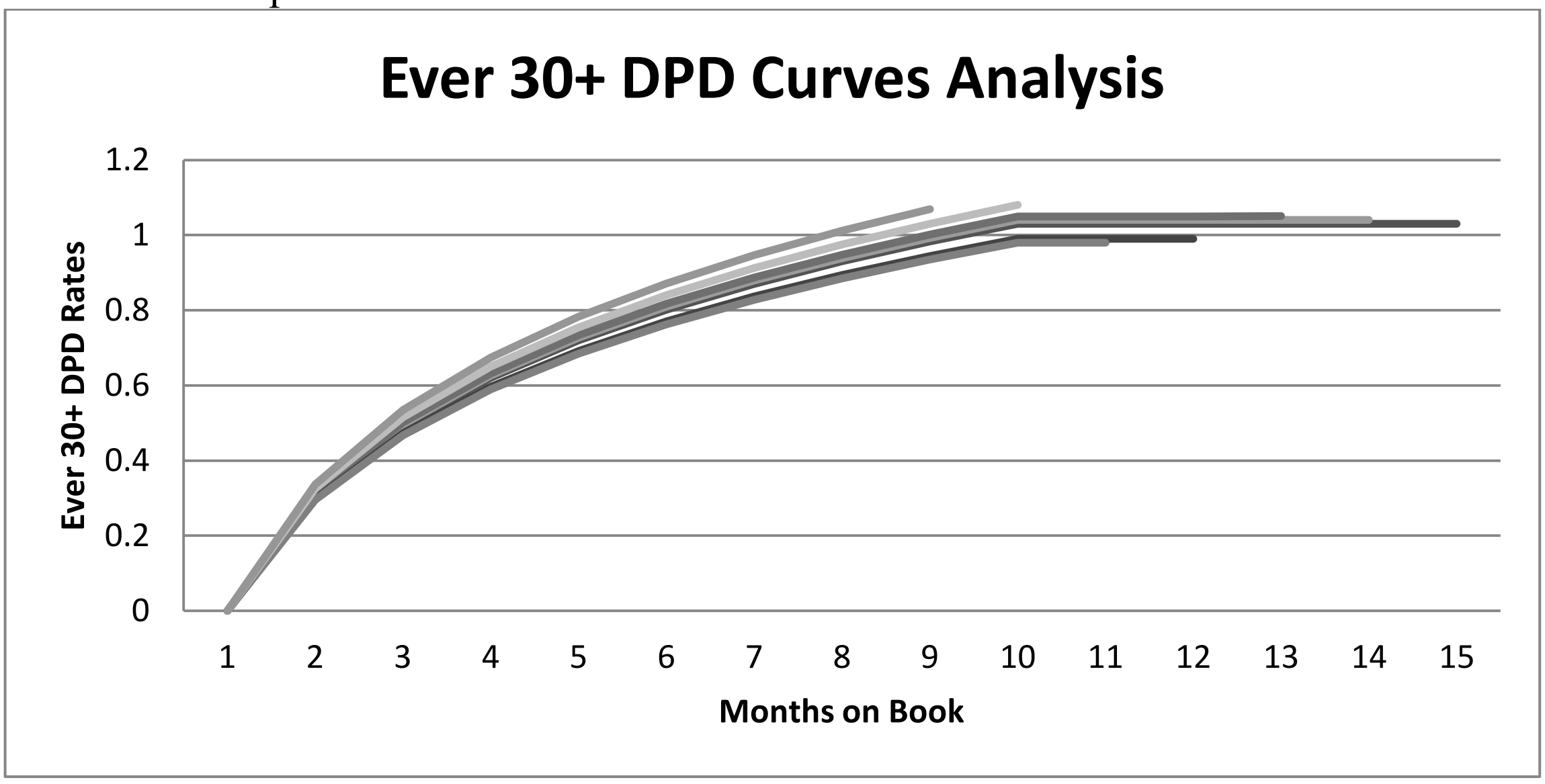

**Chart 1: Ever 30+ DPD Trend analysis (Stable leveling)**

From the chart 1, we can see the distinct flattening of the ever dpd curves. Being a simulated example, it does not capture the typical unstable flattening of the delinquency curves. Below is another simulated example that looks closer to the ones encountered by analysts in their environtment.

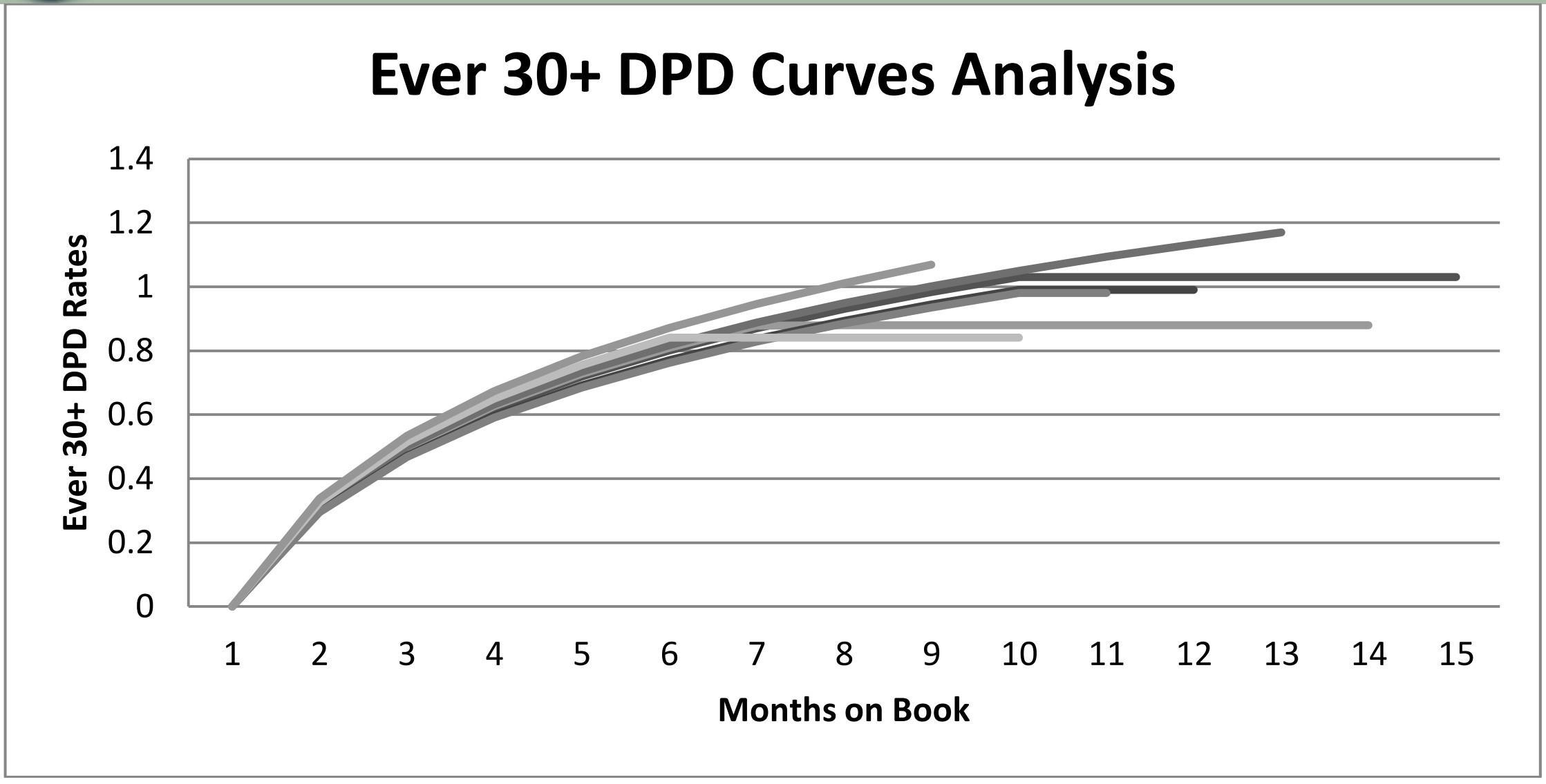

**Chart 2: Ever 30+ DPD Trend analysis (Unstable leveling)**

From chart 2, we can see that for different vintages, the flattening of the curves are differs from one another and it is extremely difficult to decide on a point in time to identify the start of the flattening. This is compounded by the problem of vintages which are almost ever increasing in their ever bad rate.

The other more serious issue with this analysis is that it requires us to preset the delinquency that will be used for the bad definition to proceed. While multiple iterations will be possible to identify the various optimum performance period for various definitions, it is ultimately a tedious and arduous process.

**ROLL RATE ANALYSIS: BAD DEFINITION ISSUES**

Once a performance period has been determined, the next important thing to set up is the delinquency definition. Most people would wonder why we do not outright use default or write off as the definition. The reason lies in the fact that outright defaults are small in number and write offs might be manipulated by the management who needs to maintain a good return and low write off portfolio. To compensate for these problems, the most common approach is to define a level of delinquency which signifies the point of no return to default. Usually, any accounts that reach a level of delinquency will have a very high chance of going straight to default. The reason for this is two folds. Firstly, any accounts which have been delinquent for a while will have accumulated massive amount of delinquent amount with interest rate compounding on them. Secondly, if the borrower wishes to repay or possesses the mean to pay, the delinquency will not slip to such a high level of delinquency.

To determine the bad definition, the traditional approach is to use the roll rate analysis (Siddiqi, 2006). Roll rate analysis is a simple Markov Model in which the accounts are grouped according to their ever delinquency status for X months and subsequently whether the account went default in the next Y months. Some variations of the technique exist and one example is the current month vs. next X month delinquency analysis (Siddiqi, 2006). Below is an example of the chart used in the analysis.

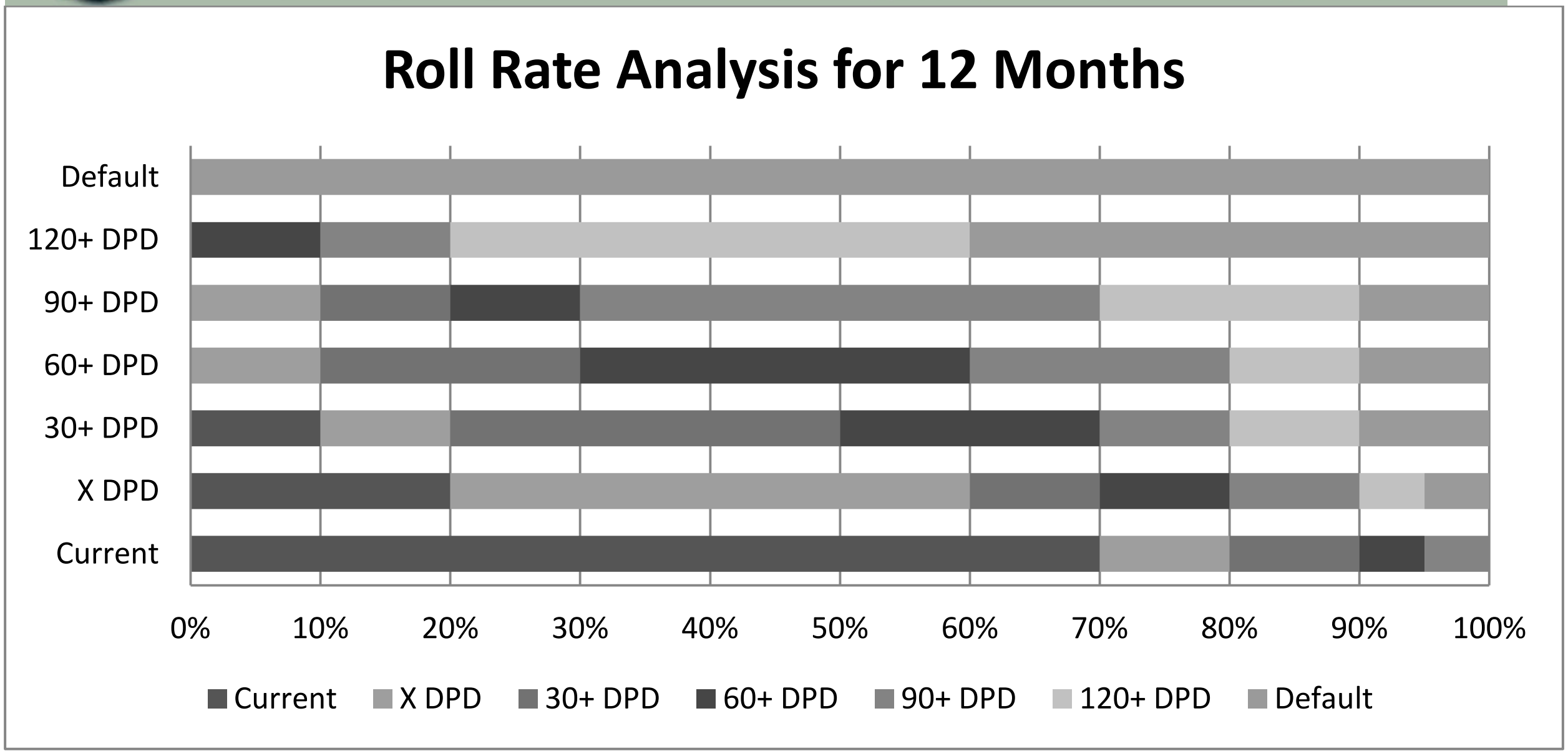

**Chart 3:Roll Rate Analysis Chart**

From chart 3, we can identify that any accounts going to 120+ DPD in first 12 months have more than 40% chance of going default in the next 12 months. This marks the delinquency status which has a huge group of people going to default once reached. However, as mentioned earlier, the difficulty in executing this analysis is the values used for X and Y in the model. Subjectivity in this case would suggest that there can be multiple definitions used for modeling and that the same chart may result in two different definitions with two analysts.

Together the traditional approaches have their good share of weaknesses which makes them undesirable. In the next section, I will introduce a more robust technique to estimate both delinquency and performance period simultaneously.

## MARKOV CHAIN: A PROPERTY THAT SOLVES THE PROBLEM

Markov Chains, also known as transition matrices, are mathematical models which define the probability of an object moving from one state to other states. Depending on the data available, there are several ways to building such a matrix. Below is the mathematical form of the matrix.

| **States** | **A1** | **A2** | **. . .** | **A(N-1)** | **A(N)** |
|---|---|---|---|---|---|
| **A(1)** | P(1,1) | P(1,2) | . . . | P(1,N-1) | P(1,N) |
| **A(2)** | P(2,1) | P(2,2) | . . . | P(2,N-1) | P(2,N) |
| **.** | . | . | . . . | . | . |
| **.** | . | . | . . . | . | . |
| **.** | . | . | . . . | . | . |
| **A(N-1)** | P(N-1,1) | P(N-1,2) | . . . | P(N-1,N-1) | P(N-1,N) |
| **A(N)** | P(N,1) | P(N,2) | . . . | P(N,N-1) | P(N,N) |

**Chart 4:Hypothetical Transition Matrix**

Each entries in the matrix represent the probability that an object will move to this state given that it starts from the state on the left per turn (usually defined as the time to transit which in this case is one month.). Total sum for each will be 1 for closed systems. One of the interesting property of the Markov chain is that one could calculate the average time spent in each transition states. This calculation is only possible in cases where the matrix contains only transient states (Referring to the case where the row summation does not total to 1). Because of this property, it happens to be uniquely qualified to solve the problem faced in solving the performance period and delinquency to default values.

Let us consider a matrix Q where the states are numbered T = {1,2,…,t} as the set of transient states.

$$\boldsymbol{Q} = \begin{bmatrix} P_{11} & P_{12} & \cdots & P_{1t} \\ P_{i1} & P_{i2} & \cdots & P_{it} \\ P_{t1} & P_{t2} & \cdots & P_{tt} \end{bmatrix}$$

For each transient state i and j, let $m_{ij}$ denote the expected total number of time periods spent in state j given the starting state of i. Reorganizing the formula yields the following result.

$$\begin{aligned} m_{ij} &= \delta(i,j) + \sum_{k} P_{ik} m_{ik} \\ &= \delta(i,j) + \sum_{k}^{t} P_{ik} m_{ik} \end{aligned}$$

Where $\delta(i,j) = 1$ when $i = j$ and 0 otherwise. Let M be the matrix containing $m_{ij}$.

$$\boldsymbol{M} = \begin{bmatrix} m_{11} & m_{12} & \cdots & m_{1t} \\ m_{i1} & m_{i2} & \cdots & m_{it} \\ m_{t1} & m_{t2} & \cdots & m_{tt} \end{bmatrix}$$

Converting it into the matrix form yields the following equation

$$M = I + QM$$

which can be transformed into

$$(I-Q)M = I$$

and with a little tweak becomes

$$M = (I-Q)^{-1}$$

## RESULTS

One important aspect of the data is the required need to filter away customers who have never been delinquent in their entire on account lifetime. This filter is needed as these accounts will artificially increase the mean time spent in current state. At the same time, as mentioned in the earlier sections, we are interested in only accounts that will go to default or write off.

Thus, only accounts that have delinquent history will be useful for us to determine the mean time spent in each state before they reached the point of no return. Let us examine the Markov chain from a credit data set after filtering as shown below.

| States | Closed | Current | X | 30 | 60 | 90 | 120+ (Write off) |
|---|---|---|---|---|---|---|---|
| Closed | 100% | 0% | 0% | 0% | 0% | 0% | 0% |
| Current | 2% | 66% | 31% | 1% | 0% | 0% | 0% |
| X | 4% | 17% | 71% | 7% | 0% | 0% | 0% |
| 30 | 4% | 4% | 15% | 45% | 30% | 3% | 0% |
| 60 | 6% | 1% | 2% | 3% | 33% | 49% | 6% |
| 90 | 3% | 2% | 1% | 1% | 2% | 26% | 66% |
| 120+ (Write off) | 0% | 0% | 0% | 0% | 0% | 0% | 100% |

**Table 2: Transition Matrix from real data**

According to the credit policy, any accounts with 120 days past due are considered as write offs. From the table, we can already observe that any accounts that start in a state of 90 DPD will have more than 50% chance to go to write off. Being the prior state before the final state, it is quite normal to have a higher rate of conversion to the next state. The 60+ DPD state also have very high conversion rate to 90 DPD as well as 120+ DPD. Comparing the conversion rate to the next state to the case of staying or moving to a better state, we can see that people who starts from 60 DPD state has less than 50% chance of staying at 60 DPD or becoming better. Given this case, we can conclude that this is the state which is the point of no return. To attempt to calculate the mean time in state, we will have to first transform the matrix into a canonical form.

| States | Current | X | 30 | 60 | 90 | 120+ | Closed |
|---|---|---|---|---|---|---|---|
| Current | 66% | 31% | 1% | 0% | 0% | 0% | 2% |
| X | 17% | 71% | 7% | 0% | 0% | 0% | 4% |
| 30 | 4% | 15% | 45% | 30% | 3% | 0% | 4% |
| 60 | 1% | 2% | 3% | 33% | 49% | 6% | 6% |
| 90 | 2% | 1% | 1% | 2% | 26% | 66% | 3% |
| 120+ | 0% | 0% | 0% | 0% | 0% | 100% | 0% |
| Closed | 0% | 0% | 0% | 0% | 0% | 0% | 100% |

**Table 3:Canonical Form of transition Matrix**

Using the transient matrix (shown in table 3 from current to 90 DPD), we can calculate the mean time for each transition state.

| States | Current | X | 30 | 60 | 90 |
|---|---|---|---|---|---|
| Current | 5.9 | 7.6 | 1.2 | 0.6 | 0.4 |
| X | 6.2 | 10.0 | 1.7 | 0.8 | 0.6 |
| 30 | 1.4 | 2.5 | 1.2 | 0.6 | 0.5 |
| 60 | 0.9 | 1.4 | 0.8 | 0.9 | 0.6 |
| 90 | 0.4 | 0.7 | 0.2 | 0.9 | 0.9 |

**Table 4: Mean Transition Time Matrix**

To calculate the mean time taken to reach 60 DPD, all we have to do is to sum the row one up to the point of 60 DPD. From table 4, the result is 5.9 + 7.6 + 1.2 + 0.6 = 15.3 which approximate to 15 months. Using this information, we can conclude that average time to reach 60 DPD is 15 months and thus the performance period can be set to 15 months
A detailed analysis of the data using the traditional techniques yielded a model with 12 months performance period and 60+ DPD as the bad definition. From this, we can see that the Markov Chain approach produces similar results in a more direct manner.

**CONCLUSION**

Markov Chain provides Credit Risk analysts with a powerful tool to define their performance period as well as the bad definition that they can use to build credit scorecards on.

***